\documentclass[pdftex,preprint, acs, amssymb, jpc,onecolumn]{revtex4}
\usepackage{amsmath}
\usepackage{amssymb}
\usepackage{amsthm}
\usepackage[pdftex]{graphicx}
\usepackage{pslatex}

\begin{document}
\title{Kerr and free-carrier ultrafast all-optical switching of GaAs/AlAs
nanostructures near the three-photon edge of GaAs}

\author{Alex Hartsuiker}\email{hartsuiker@amolf.nl}
\author{Philip Harding}
\affiliation{Center for Nanophotonics, FOM Institute for Atomic and
Molecular Physics (AMOLF), Kruislaan 407, 1098 SJ Amsterdam,
The Netherlands}%
\author{Yoanna-Reine Nowicki-Bringuier}
\author{Jean-Michel G\'erard}
\affiliation{CEA/INAC/SP2M, Nanophysics and Semiconductor
Laboratory, 17 rue des Martyrs, 38054 Grenoble Cedex, France}
\author{Willem L. Vos}
\affiliation{Center for Nanophotonics, FOM Institute for Atomic and
Molecular Physics (AMOLF), Kruislaan 407, 1098 SJ Amsterdam,
The Netherlands}%
\affiliation{Complex Photonic Systems (COPS), MESA+ Institute for
Nanotechnology, University of Twente, The Netherlands}

\date{\today, version 5}
\begin{abstract}
We performed non-degenerate pump-probe experiments on a GaAs/AlAs
photonic structure. We switched the photonic properties using the
optical Kerr effect and free carriers excited by three photon
absorption. From these measurements we extracted the non-degenerate
Kerr coefficients over a broad wavelength range and we extracted the
three photon absorption coefficient for GaAs at three wavelengths in
the near infrared. We show that the optical Kerr effect is large
enough to switch a cavity resonance with a high \emph{Q} ($Q>1000$)
photonic cavity.
\end{abstract}
\maketitle
%
\section{Introduction}

Exciting prospects arise when photonic structures are switched on
ultrafast timescales. For example, switching would allow the capture
or release of photons from photonic band gap cavities
\cite{Johnson02}, which is relevant to solid-state slow-light
schemes \cite{Yanik04}. Switching the directional properties of
photonic crystals also leads to fast changes in the reflectivity,
where interesting changes have been reported for Bragg stacks
\cite{Hache00,Hastings05}, 2D photonic crystals
\cite{Leonard02,Tan2004,Bristow03}, and first-order stop bands of 3D
opaline crystals \cite{Mazur03,Beck05}. Ultrafast control of the
propagation of light is essential to applications in active photonic
integrated circuits \cite{nakamura04}.

Different mechanisms are possible for switching photonic structures.
The switching of photonic structures with free carriers
\cite{Leonard02,Tan2004,Harding2007,Hu2008} and a phase transition
of $VO_2$ \cite{Mazur05} have been reported recently. The
disadvantage of a phase transition is that the material changes from
a transparent material to a metal, which absorbs light. On the other
hand free carrier excitation is ultrafast, but the recombination of
the carriers is limited to a picosecond time scale.

With instantaneous switching it would be possible to ultimately
control the on- and off-time of the switching event. This is
possible with the optical Kerr effect, where the on- and off-time
are determined by the pulse duration. This extreme of fast switching
can be used for capturing and releasing photons on demand for
example in a vertical-cavity surface emitting laser
\cite{Hense1997}.

In order to release photons from cavity, the refractive index change
induced with the optical Kerr effect should be large enough. To
shift the cavity resonance by one linewidth the refractive index
change should be equal to $\frac{\Delta n'}{n'} = \frac{1}{Q}$
\cite{Euseretal:08}. Thus for an experiment with a cavity having a
quality factor Q of 1000, this means a refractive index change of
0.1\%. It is well known (see \cite{Johnson02}) that the magnitude of
the refractive index change by the Kerr effect is much smaller than
the change due to a free carrier switch.

A problem which occurs at coincidence of pump and probe pulse is
non-degenerate two-photon absorption \cite{Harding2007}. Here, a
probe photon is absorbed together with a pump photon when the summed
energy of the pump and probe exceed the bandgap energy. The change
in imaginary part of the refractive index can be large compared to
the change in the real part, giving rise to absorptive changes in
the optical properties.

In this work we propose a method to decrease non-degenerate two
photon absorption at pump and probe coincidence. Because of this, we
are able to show that the optical Kerr effect is large enough to
switch a cavity resonance with a moderate Q of 1000 or higher. We
derive the Kerr coefficient and the three photon absorption
coefficient over a broad wavelength range.

\section{Experimental}
\subsection{Sample}
Our structure consists of a GaAs $\lambda$ thick layer (277 nm
thick) sandwiched between two Bragg stacks consisting of 12 and 16
pairs of $\lambda/4$ thick layers of nominally pure GaAs or AlAs.
The sample is grown with molecular beam epitaxy at 550$^{0}$C to
optimize the optical quality \cite{Gerard1996}. For experiments
outside the present scope the sample was doped with
$10^{10}$cm$^{-2}$ InGaAs/GaAs quantum dots, which hardly influence
our experiment \cite{noteQD}.

\subsection{Optical measurements}
Our setup consists of two independently tunable optical parametric
amplifiers (OPA, Topas), that are the sources of the pump and probe
beams. The setup has been described in \cite{Euser:07a} therefore we
present only a brief outline here. The OPAs have pulse durations
$\tau_{\rm{P}} = 140 \pm 10$ fs. The pump beam has a much larger
Gaussian focus of 113 $\mu$m full width at half maximum than the
probe beam (28 $\mu$m), ensuring that only the central flat part of
the pump focus is probed. A versatile measurement scheme was
developed to subtract the pump background from the probe signal, and
to compensate for possible pulse-to-pulse variations in the output
of our laser \cite{Euser:07a}. Separately, continuous-wave (cw)
reflectivity was measured with a broad band white light setup with a
resolution of $\sim 0.2$ nm \cite{Thijssen1999}.

In this work we varied the pump wavelength from 2000 nm to 2400 nm
in steps of 200 nm. The probe wavelength was varied independently of
the pump wavelength between 1150 nm and 1650 nm, with steps of 5 nm.
Measured data was corrected for dispersion in the system.

\section{Results}

\subsection{Linear reflectivity}
\begin{figure}[htb]
  \begin{center}
  \includegraphics[width=12cm]{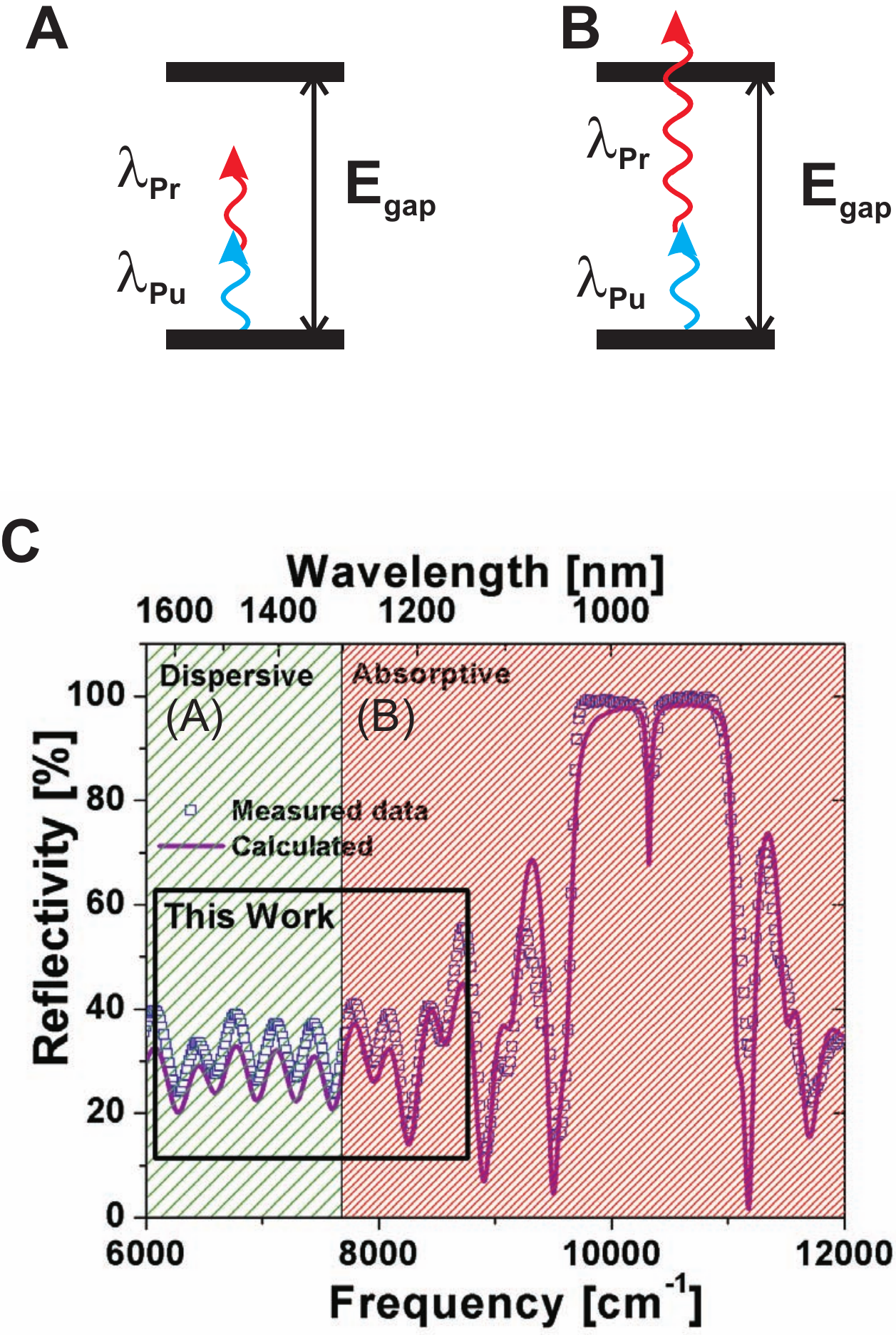}
  \caption{\emph{(A,B) Schematic energy diagrams of GaAs. $E_{gap}$ is the electronic bandgap of GaAs,
  $\lambda_{Pr}$ is the probe wavelength and $\lambda_{Pu}$ is the pump wavelength. The edge between
  diagram A and B is at a probe wavelength of 1510 nm and 1340 nm at pump wavelengths of 2000nm and
  2400nm respectively. In the dispersive spectral region the summed energy of a pump and a probe photon
  is smaller than the bandgap, while the summed energy is larger in the absorptive spectral region.
  We present measurements in the spectral region indicated with a square, to obtain a change in the real
  part of the refractive index. (C) Linear reflectivity spectrum and transfer matrix calculation of
  the GaAs/AlAs structure. The trough at 980 nm is due to the cavity resonance of the lambda thick
  GaAs layer. The hatched regions are based on a pump wavelength of 2400nm. The slight difference in
  amplitude of the measured and calculated reflectivity on the red side of the stopband is caused by a
  small error in the normalization measurement.}}\label{REflvsFreqBioRad}
  \end{center}
\end{figure}

Figure \ref{REflvsFreqBioRad}C shows the measured linear
reflectivity spectrum of our sample and a transfer matrix (TM)
calculation \cite{BornWolf}. The transfer matrix calculation
including the dispersion of GaAs \cite{Blakemore1982} and AlAs
\cite{Fern1971} reproduces the experimental resonance, stopband, and
Fabry- P\'{e}rot fringes. The only free parameters in the model were
the thicknesses of the GaAs (dGaAs = 69.2 nm) and AlAs (dAlAs = 81.0
nm), which agree to the measured values.
 The reflectivity spectrum shows
the stopband with cavity resonance and Fabry-P\'{e}rot fringes. The
spectrum can be divided into two spectral regions, namely
'Absorptive' (right, densely hatched) and 'Dispersive' (left,
hatched), referring to the expected behavior if pump and probe
coincide.

The origin of the absorption is explained in figure
\ref{REflvsFreqBioRad}A and \ref{REflvsFreqBioRad}B, which show
schematic energy diagrams at coincidence of pump and probe. The
regions are separated at a wavelength which corresponds to the
bandgap at a pump wavelength of 2400 nm. In the densely hatched
spectral region, the pump and probe photon will be absorbed since
the sum of the photon energies is higher than the bandgap of GaAs.
In the sparsely hatched (dispersive) spectral region  however, the
sum of pump and probe photon energies is not larger than the bandgap
of GaAs. Therefore absorption is low in this region and behavior at
coincidence is therefore mainly dispersive. Only the real part of
the refractive index \emph{n'} changes.

The origin of the dispersion is the optical Kerr effect. There will
be a change in the real part of the refractive index (\emph{n'}),
due to the pump field. The edge of the hatched regions in figure
\ref{REflvsFreqBioRad}C will shift to the red at a pump wavelength
of 2000 nm, indicating that we expect absorption in a large part of
the spectrum. The behavior at coincidence is therefore mainly
absorptive, meaning a change in the imaginary part of the refractive
index, (\emph{n''}).

We focus in this work on the spectral region indicated by the black
rectangle. In this region we expect to observe effects of the
optical Kerr effect but only little absorption at 2400 nm pump
wavelength, while we expect mainly absorption at 2000 nm pump
wavelength.

\subsection{Ultrafast switched reflectivity}

\begin{figure}[htb]
  \begin{center}
  \includegraphics[width=12cm]{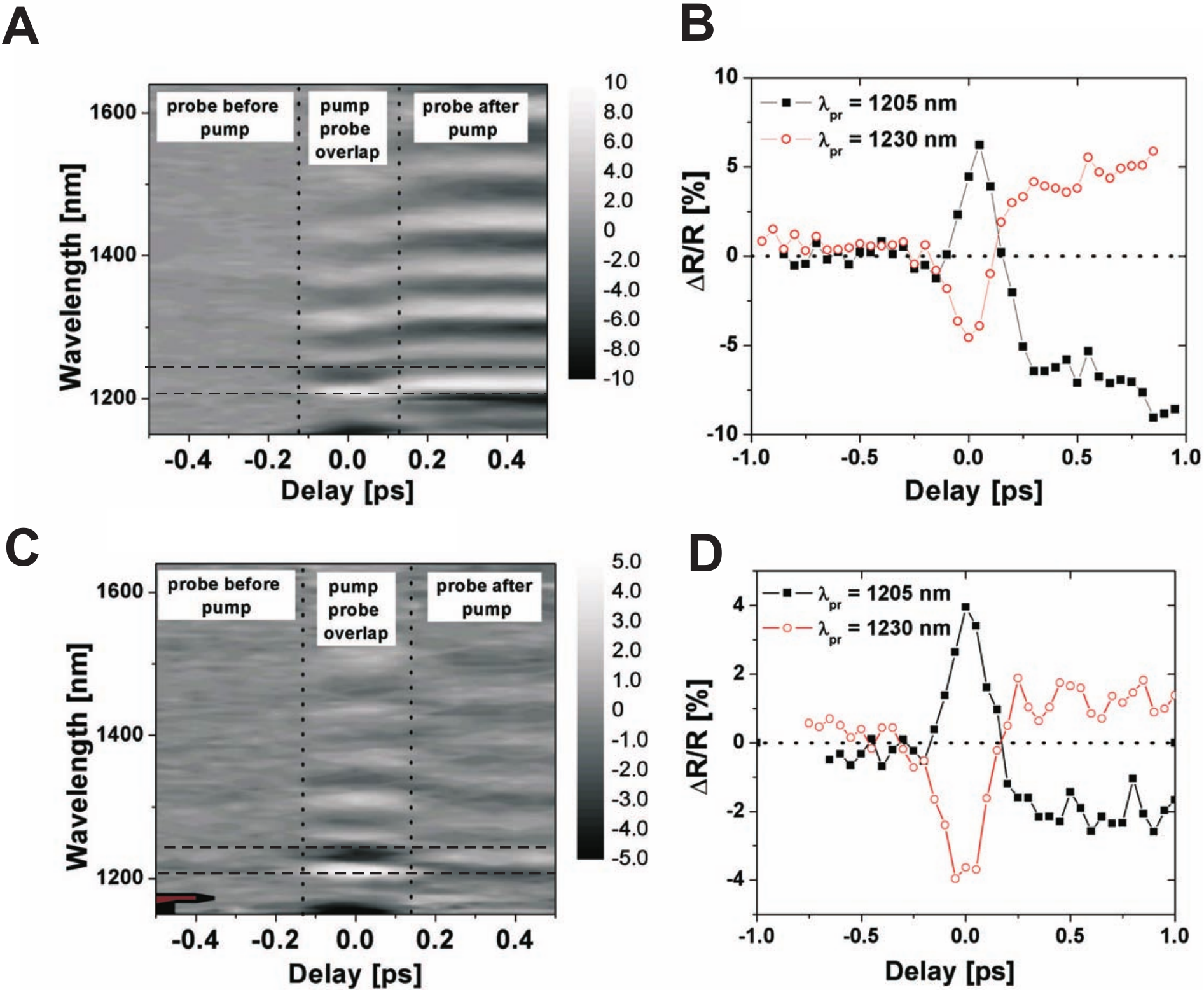}
  \caption{\emph{Differential reflectivity per wavelength as a function of delay between pump and probe
  pulse. At negative delays the pump hits the sample before the probe. The scans were measured at
  different pump wavelengths. (A) $\lambda_{pump}$ = 2000 nm, $I_{pump}$ = 90 $GW/cm^2$,
  (C) $\lambda_{pump}$ = 2400 nm, $I_{pump}$ = 95 $GW/cm^2$. Cross cuts indicated by dashed lines in A
  and C
  are shown in figure B and D respectively. Figure A and C show a fringe pattern indicating a shift of
  the Fabry-P\'{e}rot fringes. The cross sections B and D show that the sign of the differential
  reflectivity at coincidence is different from the sign at positive delay.}}\label{FullRefl}
  \end{center}
\end{figure}

The two plots in figure \ref{FullRefl} show the differential
reflectivity measured at 2000 nm, $I_{pump}$ = 90 $GW/cm^2$ (A) and
2400 nm, $I_{pump}$ = 95 $GW/cm^2$ (C) pump wavelength as a function
of probe wavelength and delay. Cross sections of figure
\ref{FullRefl}A and C are given in figure \ref{FullRefl}B and D
respectively.

A fringe pattern is visible in figure \ref{FullRefl}A and C for
coincidence of pump and probe and for positive delay. Figure
\ref{FullRefl}B and D show cross sections of \ref{FullRefl}A and C.
The sign difference of differential reflectivity between pump and
probe overlap and positive delay is apparent and is a result from
the fringe patterns at coincidence and positive delay, which are
spectrally shifted with respect to each other. This indicates that
the switch mechanism at coincidence is different from the
free-carrier mechanism at positive delay.

At a pump wavelength of 2000 nm we expect a change in \emph{n''},
since we are in the absorptive regime. At 2400 nm pump wavelength we
expect a change in \emph{n'} and a negligible change in \emph{n''},
since we are in the dispersive region (see figure
\ref{REflvsFreqBioRad}).

\begin{figure}[htb]
  \begin{center}
  \includegraphics[width=8cm]{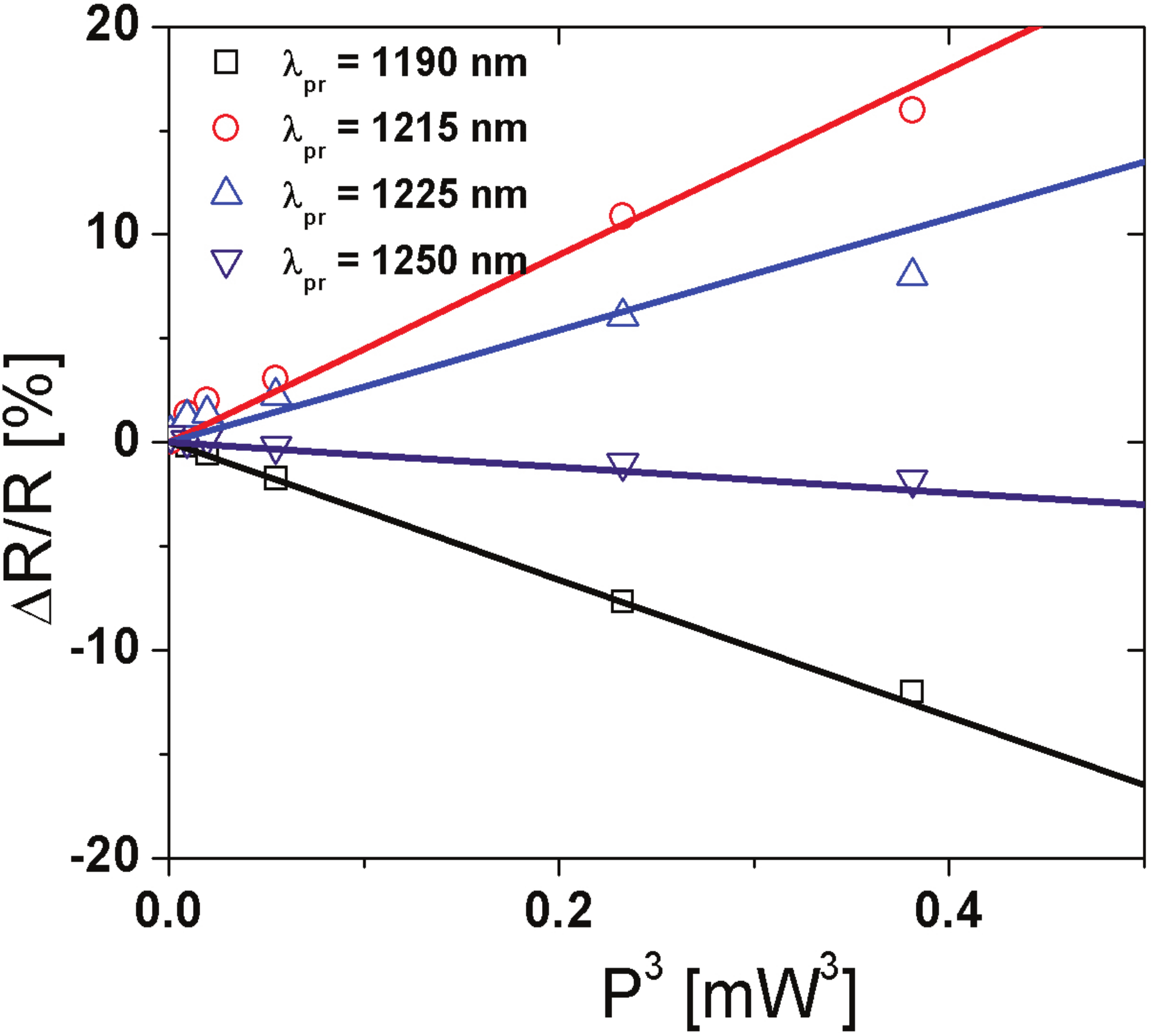}
  \caption{\emph{Differential reflectivity at positive delay measured at different probe wavelengths at
  a pump wavelength of 2000 nm. The differential reflectivity is plotted as a function of pump power cubed.
  The relation between the differential reflectivity at positive delay and the power cubed is linear.
  We conclude that the carriers are solely generated through a three photon process. }}\label{3PhotonAbs}
  \end{center}
\end{figure}

The differential reflectivity at positive delay is caused by excited
free carriers. Figure \ref{3PhotonAbs} shows the power dependence of
differential reflectivity ($\Delta R/R$) at positive delay.
There is a linear relation between the differential reflectivity and
the power cubed, as shown in figure \ref{3PhotonAbs}. We therefore
conclude that the differential reflectivity is caused by free
carriers generated by a three photon absorption process.

\subsection{Interpretation of time-resolved reflectivity}

The fringe pattern in figure \ref{FullRefl}A and \ref{FullRefl}C
results from Fabry-P\'{e}rot interferences within the stack forming
the optical microcavity. A change in \emph{n'} modifies the optical
thickness of the GaAs layers; it induces a spectral shift of the
Fabry-P\'{e}rot fringes, which results in a periodic differential
reflectivity. A change in \emph{n''} decreases the modulation depth
of the Fabry-P\'{e}rot fringes.

The differential reflectivity due to a change in the real part of
the refractive index depends on measured intensities as follows:
\begin{equation}
\frac{\Delta R}{R}(\tau, \omega_{pr},\omega_{pu}) =
\frac{I_{pu}(\tau,
\omega_{pr},\omega_{pu})-I_{unpu}(\omega_{pr})}{I_{unpu}(\omega_{pr})}
\label{eq:DRRindepth}
\end{equation}
Where $\frac{\Delta R}{R}$ is the differential reflectivity, $\tau$
is the delay between pump and probe, $\omega_{pr}$ is the probe
frequency, $\omega_{pu}$ is the pump frequency, $I_{unpu}$ is the
reflectance measured if the structure is unpumped and  $I_{pu}$ is
the reflectance measured if the structure is pumped.

\begin{figure}[htb]
  \begin{center}
  \includegraphics[width=8cm]{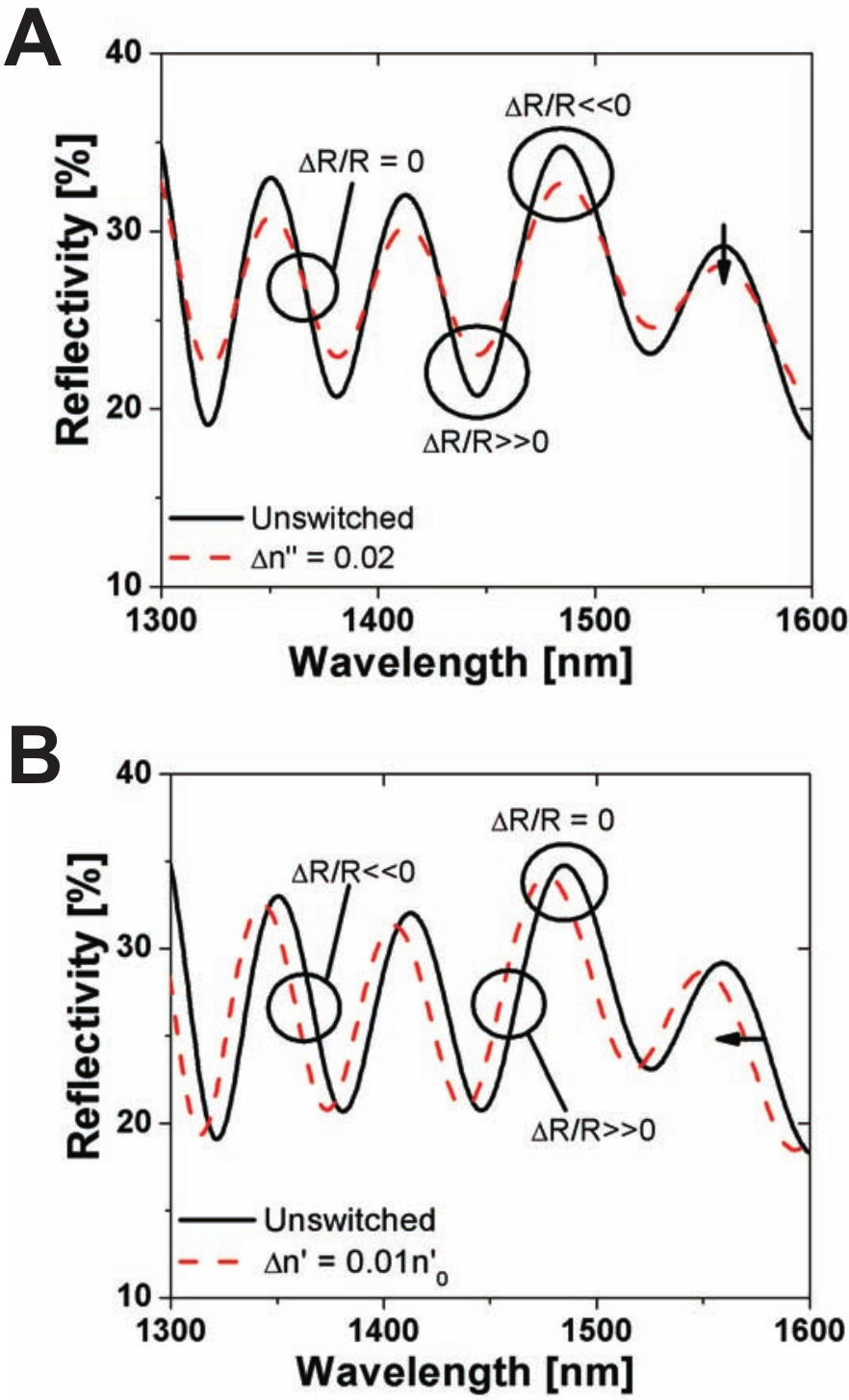}
  \caption{\emph{Switched and unswitched reflectivity for a change in the n'' (A) and a change in n' (B).
  Figure A shows that the introduction of absorption mainly affects the modulation depth of the fringes.
  A change in n' causes a shift of the fringe pattern (B). The differential reflectivity has maxima
  at different
  spectral positions, which makes it possible to distinguish between a purely dispersive and a purely
  absorptive regime.}}\label{080610ExplainDRR}
  \end{center}
\end{figure}

In case of a change in the real part of the refractive index, the
fringe pattern shifts spectrally $\Delta \omega_{pr}=\alpha n'$,
with $\alpha$ a proportionality constant that depends on the exact
structure of the fringe pattern. $I_{pu}$ is then given by:
\begin{equation}
I_{pu}(\tau,
\omega_{pr},\omega_{pu})=I_{unpu}(\omega_{pr}+\Delta\omega_{pr}(\tau,\omega_{pu}))
\label{eq:Ipudiffnreal}
\end{equation}
Which can be written as
\begin{equation}
I_{pu}(\tau,
\omega_{pr},\omega_{pu})=I_{unpu}(\omega_{pr})+\frac{\partial
I_{unpu}(\omega_{pr})}{\partial \omega_{pr}}\Delta
\omega_{pr}(\tau,\omega_{pu}), \label{eq:Ipudiffomega}
\end{equation}
which yields a differential reflectivity:
\begin{equation}
\frac{\Delta R}{R}(\tau, \omega_{pr},\omega_{pu}) =
\frac{1}{I_{unpu}(\omega_{pr})} \frac{\partial
I_{unpu}(\omega_{pr})}{\partial \omega_{pr}}\Delta \omega_{pr}(\tau,
\omega_{pu})\label{eq:DRRindepthnreal}
\end{equation}
 Equation \ref{eq:DRRindepthnreal} shows that in the case of a change in the real
part of the refractive index a large differential reflectivity will
be measured where the unpumped reflectivity has a large derivative,
which is spectrally between successive maxima and minima of the
Fabry-P\'{e}rot fringes. This is illustrated in figure
\ref{080610ExplainDRR}.
In the case of a change in \emph{n''} the modulation depth of the
fringe pattern is affected, while in the case of a change in
\emph{n'} the position of the fringe pattern is shifted. For this
reason the differential reflectivity has maxima at different
positions making it possible to distinguish between a purely
dispersive and a purely absorptive regime.

\begin{figure}[htb]
  \begin{center}
  \includegraphics[width=8cm]{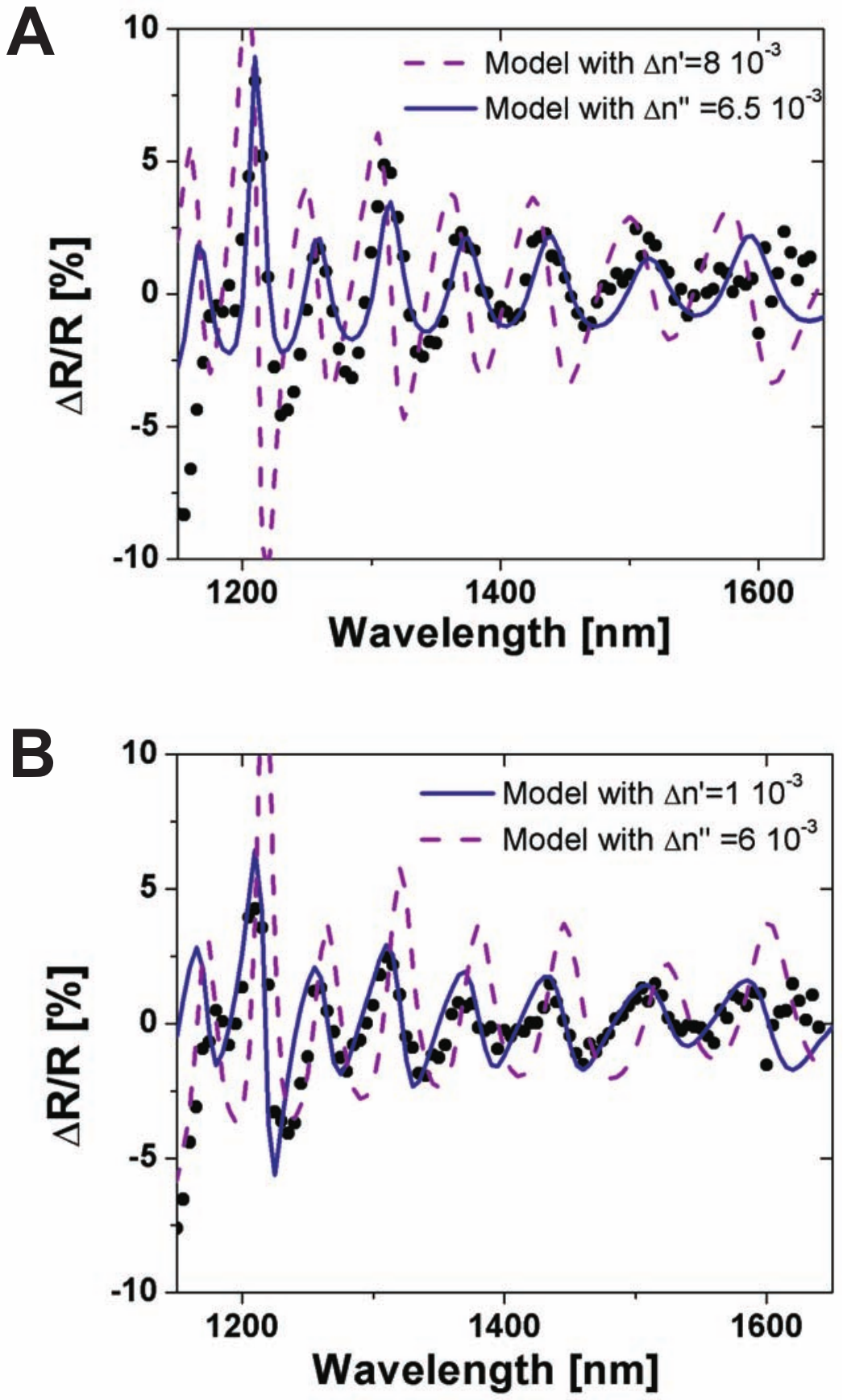}
  \caption{\emph{Cross section of figure \ref{FullRefl} (black solid circles) showing the differential
  reflectivity $\Delta R/R$ as a function of probe wavelength at pump-probe coincidence ($\Delta t=0$ ps).
  The structure
  was pumped at 2000 nm (A) and 2400 nm (B). The solid and dashed lines are results from transfer matrix
  calculations. In (A) the dashed line represents a change in the real part of the refractive index while
  the solid line represents a change in the imaginary part of the refractive index. This is the other way
  around in (B): dashed represents a change in imaginary part, while solid represents a change in the real
  part of refractive index. As expected we see mainly a change in the imaginary part of the refractive
  index at 2000 nm pump and a change in \emph{n'} at 2400 nm. Furthermore our model slightly deviates
  near the blue side of the spectrum. }}\label{DRRvsLambdaIncluReIm}
  \end{center}
\end{figure}

Figure \ref{DRRvsLambdaIncluReIm} shows a TM calculation of the
differential reflectivity spectrum of the planar microcavity. The
intensity profile in the structure is homogeneous, leading to a
homogeneous refractive index change. The intensity profile is
homogenous since we pump at a long wavelength in the three photon
absorption regime. The homogeneity length (\cite{Euser2005}) is
longer than our sample.
Figure \ref{DRRvsLambdaIncluReIm} shows a comparison between the
measured and calculated differential reflectivity pattern. Comparing
the measured and calculated patterns shows that the positions of the
peaks and troughs in the differential reflectivity pattern near zero
delay, are caused by a change in the imaginary part of the
refractive index (\emph{n''}) in the case of 2000 nm pump and a
change in the real part of the refractive index (\emph{n'}) in the
case of 2400 nm pump. We conclude that in the case of 2400 nm pump
wavelength the fringe pattern near zero delay in the differential
reflectivity originates from a Kerr switch, since the process is
ultrafast, instantaneous with the laser pulse and is dispersive. In
case of 2000 nm pump wavelength the fringe pattern originates from
non-degenerate two photon absorption. Furthermore we see a slight
deviation of our model near the blue side of the spectrum,
indicating some dispersion in the nonlinear effect. We further
conclude from figure \ref{DRRvsLambdaIncluReIm} that absorption can
be neglected at 2400 nm pump wavelength, while dispersion can be
neglected at 2000 nm pump wavelength.

\begin{figure}[htb]
  \begin{center}
  \includegraphics[width=8cm]{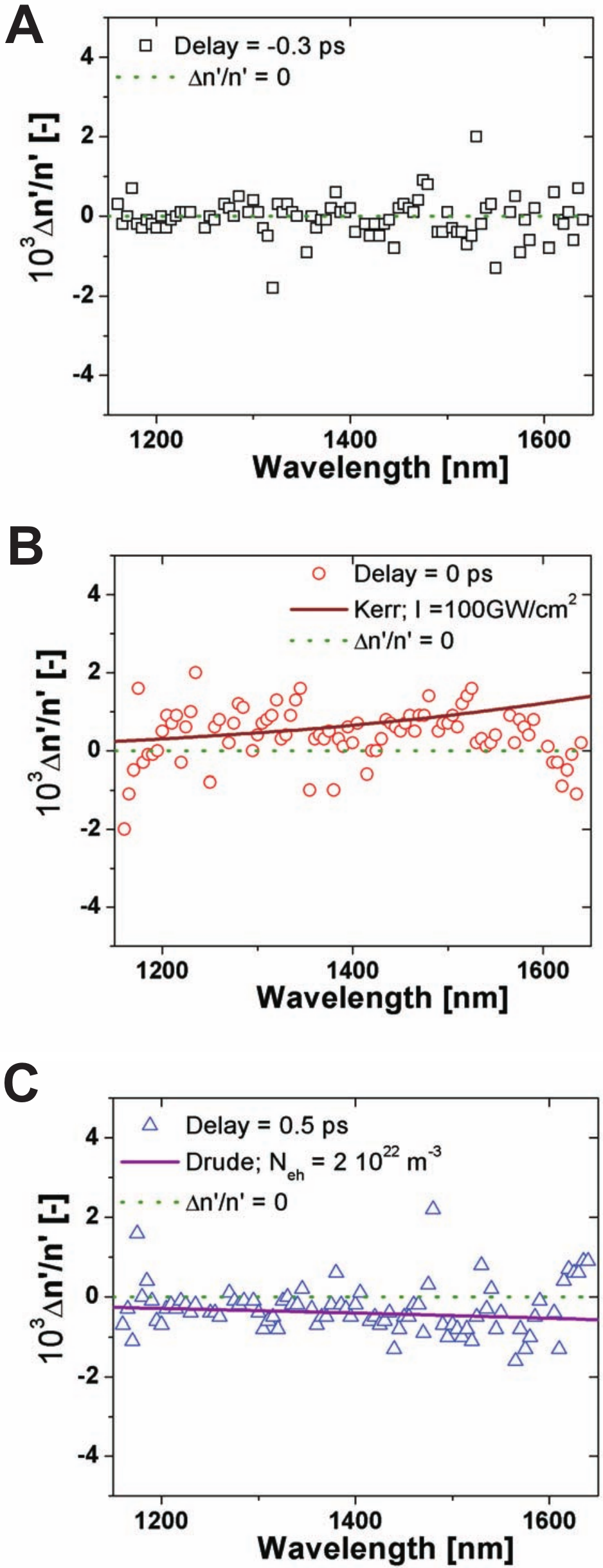}
  \caption{\emph{Relative change of refractive index due to a 2400 nm pump as a function of probe
  wavelength at a delay of -0.3 ps (A), 0 ps (B) and 0.5 ps (C). The dashed line in all three cases
  represent no change in refractive index. The solid line in (B) represents the dispersion of the change
  in refractive index from \cite{sheikbahae1994}. The solid line in (C) is calculated with the
  Drude model for free carriers \cite{Euser:07a}. Points obtained from spectral regions close to
  extrema of the fringes were removed, because of their poor precision.}}
   \label{DnnvsLambdainclKerrDrude}
  \end{center}
\end{figure}

By varying the size of the change in \emph{n'} we extracted the
change in refractive index at each wavelength. The results are
plotted together with dispersion curves in figure
\ref{DnnvsLambdainclKerrDrude}. Figure
\ref{DnnvsLambdainclKerrDrude} shows no change in refractive index
for a negative delay (A), a positive change for zero delay (B) and a
negative change for positive delay (C). Furthermore there is a good
agreement between the data and the dispersion of the optical Kerr
effect \cite{sheikbahae1994} and the dispersion of the free carrier
excitation \cite{Euser:07a}.

The scattered symbols deviating strongly from the theoretical model
can be attributed to small differences in shape of the measured and
calculated reflectivity spectrum. A slight deviation is amplified by
the fitting procedure since we extract the refractive index change
by fitting the differential reflectivity at one wavelength position.

Figure \ref{DnnvsLambdainclKerrDrude}B shows that the change in
refractive index induced by the optical Kerr effect is in the order
of 0.1\%. This is large enough to switch a cavity with a moderate Q
of 1000.

\subsection{Non linear coefficients
GaAs}\label{sec:nonlinearcoefficients}

\subsubsection{Kerr coefficient $n_{2}$ for GaAs}
The non-degenerate Kerr coefficient $n_{2}$ can directly be
extracted from the data $\Delta n'$ in figure
\ref{DnnvsLambdainclKerrDrude} using the relation \cite{Boyd,
sheikbahae1994},
\begin{equation}
n_{2}=\frac{\Delta n'}{2I_0} \label{eq:derkerrcoefficient}
\end{equation}
where $I_{0}$ is the pump intensity. Since $n_2$ scales with
$E_{gap}^{-4}$ \cite{sheikbahae1994}, the contribution of AlAs is
only 20\% of the total $n_2$. For simplicity we therefore assume
that mainly the GaAs is switched.

\begin{figure}[htb]
  \begin{center}
  \includegraphics[width=12cm]{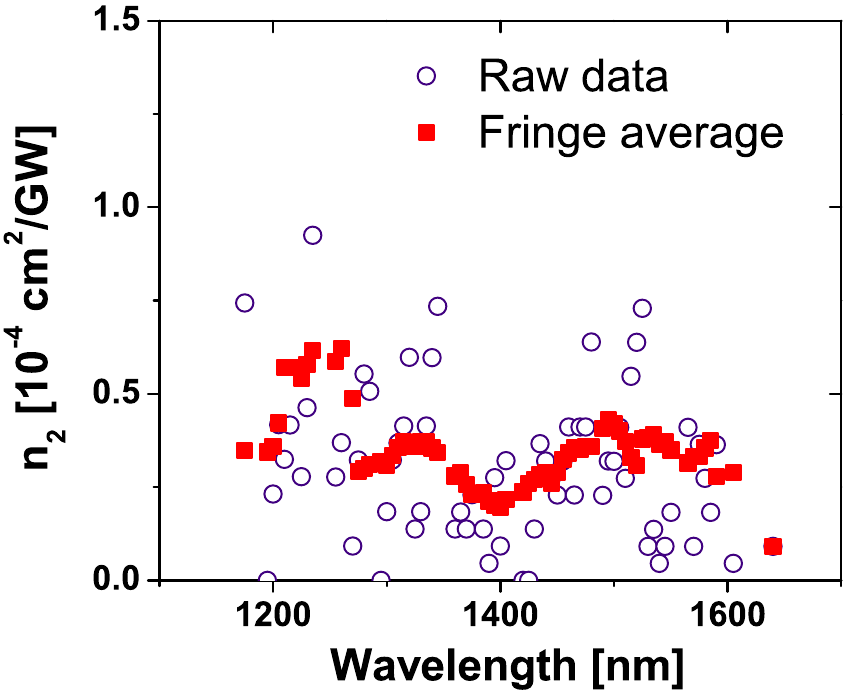}
  \caption{\emph{Measured non-degenerate Kerr coefficient $n_2$ as a function of probe wavelength
  (open circles). We averaged the data over the period of one fringe, since the coefficients are
  correlated within this fringe period (solid squares). We observe dispersion in $n_2$ towards the
  blue side of the spectrum as expected from figure \ref{DRRvsLambdaIncluReIm}. }}\label{n2vsProbePump}
  \end{center}
\end{figure}

 The resulting Kerr coefficients are plotted in figure
\ref{n2vsProbePump}.
Figure \ref{n2vsProbePump} shows the raw data extracted from figure
\ref{DnnvsLambdainclKerrDrude} (open circles) and the data averaged
over the width of a fringe (solid squares). The order of magnitude
of $n_2$ is $10^{-4} GWcm^{-2}$. The non-degenerate Kerr coefficient
decreases with increasing wavelength. This dispersive behavior was
already observed in figure \ref{DRRvsLambdaIncluReIm}.

The values of $n_2$ measured in our non-degenerate pump-probe
experiment are similar to the ones reported in the literature for
degenerate pump-probe experiments
\cite{Hurlbut2007,Dinu2003,Ulmer1999}. We find this surprising since
we pump at a wavelength twice as long as in the degenerate case.
This result enables us to modify efficiently the refractive index
with a far detuned pump wavelength.

\subsubsection{Three photon absorption coefficient $\gamma$ for GaAs}
Figure \ref{DnnvsLambdainclKerrDrude} shows that we are able to
derive a carrier density from the measured change in refractive
index with the use of the Drude model. We will discuss in this
section the three photon absorption coefficient $\gamma$.

\begin{figure}[htb]
  \begin{center}
  \includegraphics[width=8cm]{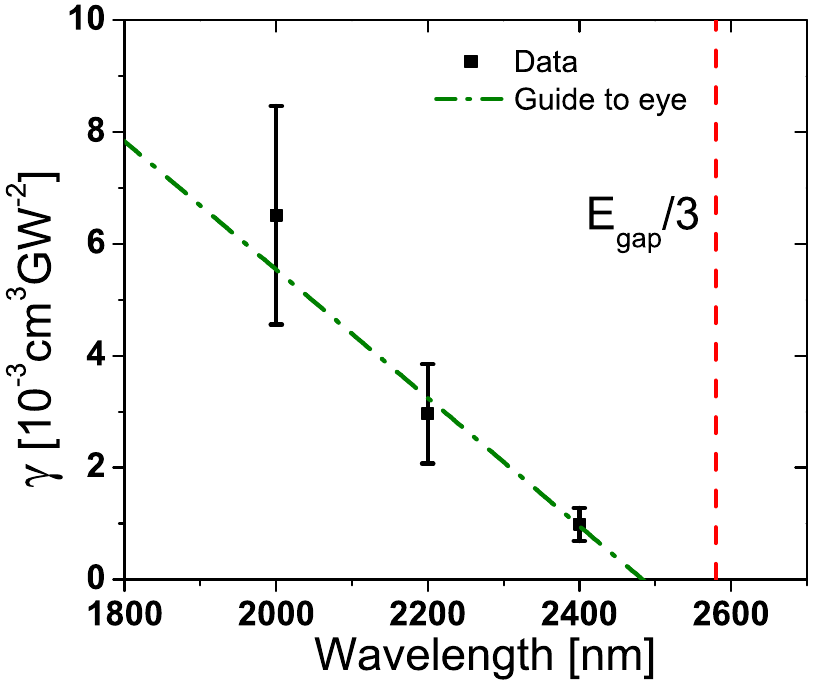}
  \caption{\emph{Three photon absorption coefficient as a function of wavelength extracted from
  the differential reflectivity data. The relative error of 30 \% is indicated. }}\label{gammavsWL}
  \end{center}
\end{figure}
Figure \ref{gammavsWL} shows the three photon absorption coefficient
$\gamma$ as a function of pump wavelength. The order of magnitude of
$\gamma$ is $10^{-3}cm^3GW^{-2}$. Our data are two orders of
magnitude smaller than the values reported in the literature
\cite{Hurlbut2007}, measured with the z-scan method. We attribute
this difference to the fact that our data have been obtained on an
epitaxially-grown GaAs/AlAs heterostructure, instead of a GaAs wafer
obtained with a different technique.

The three photon absorption coefficient decreases as a function of
wavelength since the summed energy of a pump and probe photon, and
therefore the probability of generating an electron hole pair,
decreases with increasing wavelength. The three photon absorption
edge is 2580 nm.

\section{Conclusion}

We showed switching of the fringes of a GaAs/AlAs planar optical
microcavity using the optical Kerr effect and three photon
absorption.

From the switching measurements we extracted the non-degenerate Kerr
coefficients over a broad wavelength range. Furthermore we extracted
the three photon absorption coefficients for three different
wavelengths in the near infrared.

We conclude from our results that it is possible to switch a cavity
with a Q higher than 1000, using the optical Kerr effect. The
refractive index change is in the order of the required 0.1\%.

\section*{Acknowledgments} \label{sec:Acknowledgement}
We want to thank Allard Mosk and Patrick Johnson for stimulating
discussion. This research was supported by NanoNed, a nanotechnology
programme of the Dutch Ministry of Economic Affairs, and by a VICI
fellowship from the "Nederlandse Organisatie voor Wetenschappelijk
Onderzoek" (NWO) to WLV. This work is also part of the research
programme of the "Stichting voor Fundamenteel Onderzoek der Materie"
(FOM), which is financially supported by the NWO.


\begin{thebibliography}{}
\bibitem{noteQD}
The maximum unbroadened refractive index change of the dots amounts
to only $10^{-8}$, while the absorption at resonance is less than
$0.02$ cm$^{-1}$.
\end{thebibliography}
\end{document}